\documentclass{ws-procs9x6}

\begin{document}

\title{Gamma Ray Bursts as Cosmological Probes\footnote{\uppercase{T}alk presented 
by one of us (\uppercase{P.T.S.}) at the \uppercase{XV} \uppercase{E}ncontro \uppercase{N}acional de 
\uppercase{A}stronomia e \uppercase{A}strof\'\i sica, \uppercase{L}isbon,
\uppercase{P}ortugal, 28-30 \uppercase{J}uly 2005.}}

\author{O. Bertolami and P.T. Silva}

\address{Instituto Superior T\'ecnico, Departamento de F\'\i sica\\
Av. Rovisco Pais, 1049-001\\ 
Lisboa, Portugal\\ 
E-mail: orfeu@cosmos.ist.utl.pt; paptms@ist.utl.pt
}

\maketitle

\abstracts{
We discuss the prospects of using Gamma Ray Bursts (GRBs) as
high-redshift distance estimators, and consider their use in the study of two
dark energy models, the Generalized Chaplygin Gas (GCG), a model for the unification 
of dark energy and dark matter, and the XCDM model, a model where a generic dark
energy fluid like component is described by the equation of state, 
$p= \omega \rho$. Given that the GRBs range of redshifts is rather high, 
it turns out that they are not very sensitive to the dark energy component, being however, 
fairly good estimators of the amount of dark matter in the Universe.
}

\section{Introduction}

Recently, there has been a flurry of activity about the prospect
use of GRBs as cosmological probes\cite{bertolami06}. In the original 
proposal\cite{schaefer03}, it has been suggested that
the magnitude versus redshift plot, could be extended
to a redshift up to $z\simeq 4.5$ via correlations found
between the isotropic equivalent luminosity, $L_{iso}$, and two GRB
observables, namely the time lag ($\tau_{lag}$) and variability
($V$). The isotropic equivalent luminosity is the inferred luminosity (energy
emitted per unit time) of a GRB if all its energy is radiated  isotropically,
the time lag measures the time offset between high- and low-energy arriving
GRB photons, while the variability is a measure of the
complexity of the GRB light curve. Unfortunately, these correlations are
affected by a large statistical (or intrinsic) scatter. This statistical
spread affects not only the cosmological precision via its direct statistical
contribution to the distance modulus uncertainty, $\sigma_\mu$, but also through 
the calibration uncertainty given that the suitable GRB sample with known redshift is
rather small. 
In what follows we show that a relatively small sample of GRBs with low redshifts is
sufficient to greatly reduce the systematic uncertainty thanks to a more robust
and precise calibration.

More recently, a new correlation was suggested\cite{ghirlanda04a}, which is
subjected to a much smaller statistical scatter. 
The so-called Ghirlanda relation, is a correlation between the
peak energy of the gamma-ray spectrum, $E_{peak}$ (in the $\nu - \nu F_{nu}$
plot), and the corrected collimation energy emitted in gamma-rays,
$E_\gamma$. This collimation energy is a measure of the energy
released by a GRB taking into account that the energy is beamed into a 
narrow jet. Unlike the $L_{iso}-\tau$ and $L_{iso}-V$ relations, the
Ghirlanda relation is not affected by large statistical
uncertainties, but is dependent on poorly constrained quantities related to the
properties of the medium around the burst.

Another difficulty involving GRBs is that they tend to occur at rather large distances,
which makes it impossible to calibrate any relationship between the relevant variables 
in a way that is independent from the cosmological model. The method that is 
usually employed consist in fitting both, the cosmological \emph{and} the 
calibration parameters, and then use statistical techniques to remove the
undesired parameters. In here, we follow a different procedure\cite{bertolami06,takahashi}. 
We consider a luminosity distance for $z<1.5$ that is measured by SNe Ia,
and divide the GRBs sample in two sets; the low redshift sample, with 
$z<1.5$, and the high redshift one, with $z>1.5$.
Since the luminosity distance of GRBs in the range $z<1.5$ is now known, one can 
calibrate the luminosity estimators independently of the cosmological
parameters and use the high redshift sample as a probe to dark energy and dark matter
models.

We have analyzed the use of these several correlations in the study of the GCG, a
model that unifies the dark energy and dark matter in a single fluid
\cite{bento02} through the equation of state 
$ p_{ch} = - A / \rho_{ch}^\alpha$, where $A$ and $\alpha$ are positive
constants. The case $\alpha=1$ describes the 
the Chaplygin gas, that arises in different theoretical scenarios.
If the total amount of matter is fixed, there are only two free variables,
$A$ and $\alpha$, although it is more customary to use the quantity
$A_s \equiv A/ \rho_{ch,0}^{1+\alpha}$ instead of $A$. Thus, we consider
two free parameters, $\alpha$ and $A_s$. A great deal of effort has been 
recently devoted to  constrain the GCG model parameters\cite{bertolami05}, which 
include, for instance, gravitational lensing\cite{silva03} and cosmic
topology\cite{bento05a}.

In addition to the GCG model, we also study the more conventional
flat XCDM model. Likewise the GCG model, the
XCDM model is also described by two free parameters, the parameter, $\omega$,
of the  dark energy equation of state $p = \omega \rho$, and the fraction of
of dark matter, $\Omega_m$. The test of these models is particularly
interesting since it is known that they are degenerate for redshifts 
$z<1$~\cite{bertolami04,bento05b}.

\section{Variability and Time Lag as Luminosity Estimators}

Our work can be divided in two parts. First, we test the calibration
procedure. The small sample of GRBs with measured redshifts means that at
present the calibration is rather poor. We assess the gain in 
calibrating the relations with larger samples by generating 
three mock samples and by performing their calibration.
We find that a calibration done with $40$ GRBs will greatly improve the
previous results, decreasing $\sigma_\mu$,
by close to half, yielding $\sigma_\mu=0.68$. However, by increasing the 
calibration sample to $100$ GRBs the resulting decrease is just marginal, suggesting
that is not much of a use to consider very large calibration samples. Its worth noting
that a sample of about $40$ GRBs may be available in the near future, thanks
to the {\it Swift} satellite. We also find that despite the large statistical
scatter, thanks to the improved calibration, the uncertainty for this estimator
becomes quite close to that of the Ghirlanda relation, that is, $\sigma_\mu = 0.5$.

\begin{figure}[ht]
\centerline{\epsfxsize=4.1in\epsfbox{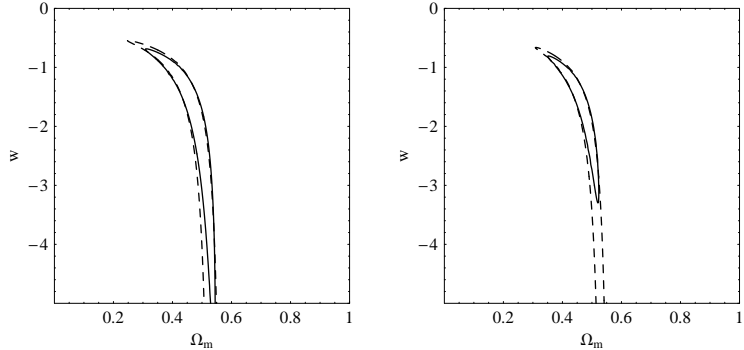}}   
\caption{Confidence regions found for the XCDM model. The solid lines show
the 68\% CL regions obtained through a sample of 100 low-redshift ($z<1.5$)
and 400 high-redshift ($z>1.5$) GRBs, while the dashed lines show the 68\% CL
constraints for a sample made up of 500 high-redshift GRBs only. On the left
figure, the $\tau-L_{iso}$ and $V-L_{iso}$ relations have been used, while on the right one
the Ghirlanda relation was employed. \label{xcdm}}
\end{figure}

We have then examined how GRBs fared when used to constrain both models
under consideration. Somewhat against our
expectation, we found that GRBs are not very suited to study the GCG model. The
results for the XCDM model are more promising, however we find that GRBs are
sensitive essentially to $\Omega_m$, and very weakly sensitive to $\omega$.
The reason for these results is the redshift range where GRBs lay. We have verified
that when using a sample that includes 100 with $z<1.5$ GRBs, the constraints
on the XCDM model where much better, as is depicted in Fig. \ref{xcdm}. 
These results were found using the minimal $\sigma_\mu=0.66$. This
uncertainty is essentially due to the statistical component, and hence it 
cannot be reduced by better calibration or data.

We also tested the use of the Ghirlanda relations, which is intrinsically more precise.
As before, we find that characteristic feature of GRBs of having rather high redshifts,
make them somewhat unsuitable to study dark energy models, even the GCG one. One finds that
results are better if one uses the Ghirlanda relations, but in what concerns dark energy
models not crucially. It should be noted that data quality and statistics will greatly
improve in the future thanks to {\it Swift} and {\it HETE 2} experiments.

Nevertheless, our main conclusion is that although GRBs are poor dark energy probes,  
for $z>1.5$, the luminosity distance is quite sensitive to the dominating 
component at the time. For the XCDM model, this is dark matter, and we find
that the amount of dark matter can be remarkably constrained. For the GCG
model, on the other hand, it turns out that what arises is a combination of
$A_s$ and $\alpha$ parameters, and data cannot lift the degeneracy on $\alpha$.
Actually, this feature is encountered in various phenomenological  studies of
the GCG, the only exception being on data from large scale structure
formation\cite{bento04}.

\end{document}